\documentclass[aps,pra,reprint,superscriptaddress,amsmath,amssymb]{revtex4-2}
\usepackage{graphicx}
\usepackage{xcolor}
\usepackage{tabularx}
\usepackage[colorlinks=true,bookmarks=false,citecolor=blue,urlcolor=blue,allcolors=blue]{hyperref} 

\usepackage{ulem} 

\begin{document}

\title{Beam propagation in an active nonlinear graded-index fiber}
\author{Anuj P. Lara}
\affiliation{Department of Physics, Indian Institute of Technology Kharagpur, West Bengal 721302, India}
\author{Samudra Roy}
\email{samudra.roy@phy.iitkgp.ac.in}
\affiliation{Department of Physics, Indian Institute of Technology Kharagpur, West Bengal 721302, India}
\author{Govind P. Agrawal}
\affiliation{The Institute of Optics, University of Rochester, Rochester, New York 14627, USA}

\begin{abstract}
A theoretical model is developed by exploiting the variational technique to investigate the evolution of an optical beam inside an optically pumped graded-index fiber amplifier. The variational analysis is a semi-analytical method that provides us with a set of coupled ordinary differential equations for the beam's four parameters. Numerical solution of these equations is much faster compared to the underlying multidimensional nonlinear wave equation. We compare the results of the variational and full numerical simulations for the two pumping schemes used commonly for high-power fiber amplifiers. In the clad-pumping scheme, the use of a relatively wide pump beam results in a nearly uniform gain all along the fiber. In the case of edge pumping, a narrower pump beam provides gain that varies both radially and axially along the fiber's length. In both cases, the variational results are found to be in good agreement with time-consuming full numerical simulations. We also derive a single equation for the beam's width that can predict amplification-induced narrowing of the signal beam in most cases of practical interest.

\end{abstract}
\maketitle

\section{Introduction}

In recent years, multimode graded-index (GRIN) fibers have been used for studying a variety of intriguing nonlinear effects such as the formation of multimode solitons \cite{Renninger_Optical_2013}, creation of dispersive waves \cite{Wright_Ultrabroadband_2015, Wright_Controllable_2015} over a wide spectral range, spatiotemporal mode-locking of lasers \cite{Wright_Self_2016, Wright_Spatiotemporal_2017, Qin_Observation_2018}, and supercontinuum generation \cite{LopezGalmiche_Visible_2016,Eslami_Two_2022}.

Multimode fibers are also useful for high-power applications because they contain a relatively wide central core. For this reason, GRIN fibers have become an obvious choice for making high-power amplifiers and lasers. Such devices provide better beam quality compared to the step-index fibers, owing to the phenomena of spatial beam cleanup \cite{Guenard_Kerr_2017, Jima_Numerical_2022}. A mode-based analysis of spatial beam cleanup shows that a GRIN fiber is instrumental in improving the beam's quality \cite{Mangini_Statistical_2022,Haig_Gain_2023}. However, this approach becomes less useful when many modes of the GRIN fiber are excited by the pump and signal beams. Non-modal numerical studies have also been carried out in recent years \cite{Sidelnikov_Mechanism_2022}. Even though a detailed numerical analysis may be needed in some situations, it requires solutions of coupled multidimensional partial differential equations and is, by necessity, resource intensive.

In this work we adopt a variational approach to model the evolution of an optical beam as it is amplified inside an active GRIN fiber, doped uniformly to provide gain. An approximate analytical treatment used recently \cite{Agrawal_Spatial_2023} ignored an important nonlinear effect known as self-phase modulation (SPM). We not only include SPM but also consider a more realistic model for the gain distributed along the GRIN fiber. Further, we consider both the edge- and side-pumping schemes and compare them. The proposed semi-analytical treatment based on the variational method is less time consuming compared to full numerical simulations and also provides considerable physical insight.

The article is organized as follows. In Section II we outline the basic theory of a an optical beam's amplification inside an active GRIN fiber and discuss the simplifications that we make to obtain a multidimensional partial differential propagation equation satisfied by the slowly varying amplitude of the signal beam. This equation is solved approximately in Section III using the semi-analytical variational approach. A suitable Lagrangian is used with the Gaussian ansatz to obtain four coupled ordinary differential equations describing the evolution of four beam parameters  inside the GRIN fiber. These equations are solved in Section IV for two specific pumping schemes and the results are compared with those obtained by solving the multidimensional equation numerically. We summarize our main conclusions In Section V.

\section{Theory}

\begin{figure}[tb!]
	\includegraphics[width=\linewidth ]{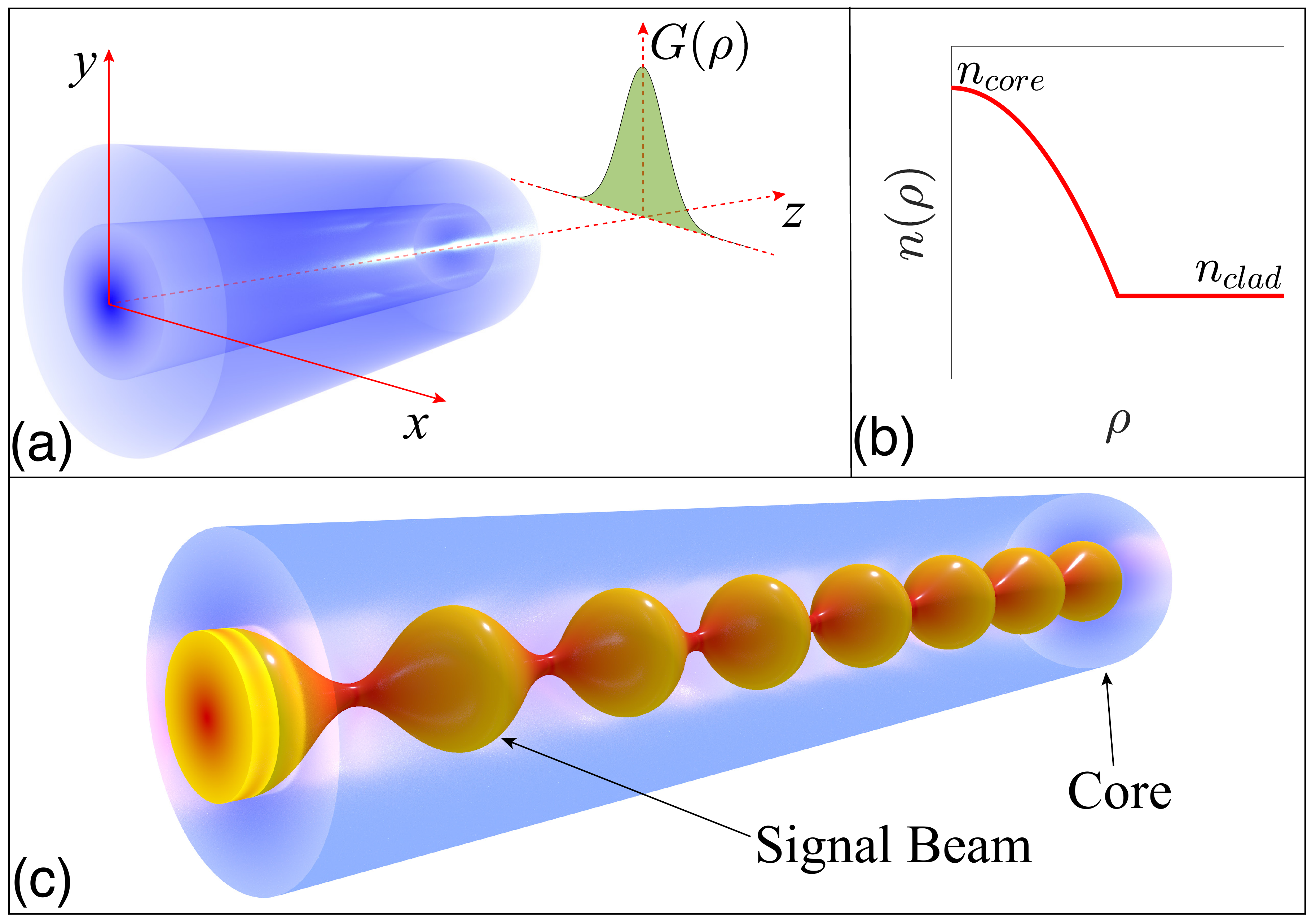}
\caption{\textbf{} Schematic of a GRIN fiber amplifier having (a) a radially varying gain and \textbf{(b)} a parabolic index profile. \textbf{(c)} Schematic showing how the signal beam evolves in a periodic fashion inside such a  GRIN fiber because of self-imaging provided by the parabolic index profile.} \label{Fig1}
\end{figure}

We consider a GRIN fiber with a parabolic index profile (see Figure \ref{Fig1}(b)) and a uniform density of dopants  along the radial direction. When such a fiber is pumped with a high-power laser, the gain can still vary radially and axially because of nonuniform inversion of the dopants, as shown schematically in Figure \ref{Fig1}(a). Two pumping schemes are used in practice. In the side-pumping scheme, a wide pump beam is launched into a double-clad fiber, resulting in a nearly uniform gain all along the fiber. In the case of edge pumping, a narrower pump beam is launched at the front end of the amplifier, resulting in a local gain $G(\rho,z)$ varying both radially and axially all along the fiber's length.

In this work, we consider both pumping schemes by including the local gain $G(\rho,z)$ through the imaginary part of the refractive index. The real part of the refractive index includes the parabolic radial  variations together with Kerr nonlinearity. The resulting expression for the refractive index becomes
\begin{equation}
    n(\rho,z) = n_{\rm core} \left(1-\frac{1}{2}b^2 \rho^2\right)+n_2|\mathbf{E}_s|^2 -i\frac{G(\rho,z)}{2n_{core}k_0},
    \label{Eq1}
\end{equation}
where $\rho = \sqrt{x^2+y^2}$ is the radial distance from the centner of the GRIN fiber and $n_{\rm core}$ is the refractive index at $\rho =0$ [see Fig.~\ref{Fig1}(b)]. The index gradient $b$ is defined as $b = \sqrt{2\Delta}/a$, where $a$ is the core's radius and $\Delta$ is the relative core--cladding index difference defined as $\Delta=1-n_{\rm clad}/n_{\rm core}$. The Kerr coefficient $n_2$ and has a value of $2.7 \times 10^{-20}$ m$^2$/W for silica fibers. The gain $G(\rho,z)$ depends on the local density of dopants and in general varies both with $\rho$ and $z$. The signal to be amplified is taken in the form of a quasi-continuous beam with a narrow spectrum centered at $\omega_0$. The wave number $k_0$ in \ref{Eq1} is defined at this frequency as $k_0= \omega_0/c$.

The electric field associated with the signal beam satisfies the \textit{Helmholtz equation}:
\begin{equation}
    \mathbf{\nabla}^2\mathbf{E}_s+n^2(\rho,z)k_0^2 \mathbf{E}_s=0.
    \label{eq2}
\end{equation}
By writing $\mathbf{E}_s$ in the form $ \mathbf{E}_s=\hat{\mathbf{p}}A_s(\rho,z)e^{ik_sz}$, where $\hat{\mathbf{p}}$ is the polarization unit vector, $k_s=n_{\rm core}k_0$, and $A_s(\rho,z)$ is the slowly varying amplitude of the signal. A paraxial approximation then leads to the following equation for the slowly varying amplitude:
\begin{align}
	i\frac{\partial A_s}{\partial z} + \frac{\nabla_{\bot}^2 A_s}{2k_s} 
    - \frac{1}{2}k_s b^2 \rho^2 A_s + \frac{\omega_0}{c}n_2|A_s|^2A_s= \notag \\
	\frac{i}{2}G(\rho,z)A_s - \frac{i}{2}\alpha_{l}A_s
    \label{eq3}
 \end{align}
where $\nabla_{\bot}^2 = \partial^2/\partial \rho^2 + \frac{1}{\rho}\partial/\partial\rho$ is the transverse Laplacian and $\alpha_l$ represents linear loss for the signal wave. In the preceding equation, the effects of diffraction, index gradient, and self-phase modulation (SPM) are included through the second, third and fourth terms, respectively. 

Before solving Eq.~\eqref{eq3}, an explicit form of the gain function $G(\rho,z)$ should be specified. In general, $G(\rho,z)$ is found by solving a set of rate equations for a given intensity profile of the pump beam, whose absorption leads to population inversion \cite{Saleh_Modeling_1990,Jima_Numerical_2022,Chen:23}. Although such an approach may be necessary for fitting the experimental data, our objective here is to develop a semi-analytical model of beam's amplification that contains essential physics of the problem with reasonable accuracy. We expect the gain to follow the radial shape of the pump beam, which we assume to be Gaussian and write the gain function $G(\rho,z)$ in the form
\begin{equation}
	G(\rho,z)=G_0(z)\exp\left(-\frac{\rho^2}{\rho_g^2}\right),
	\label{eq4}
\end{equation}
where the peak gain $G_0(z)$ (in units of m$^{-1}$) can vary with distance and the parameter $\rho_g$ is related to the pump beam's spot size.

\section{Variational analysis}

Using the explicit form of the gain profile in Eq.~(\ref{eq4}), we can solve numerically Eq.\ (\ref{eq3}) to simulate amplification of the signal beam. However, numerical simulations are found to be quite time-consuming for fibers longer than a few meters. A numerical approach also hinders physical insight and does not reveal what parameters are most relevant for narrowing of the signal beam to occur. For these reasons, we adopt the variational method \cite{Anderson_Variational_1983} for solving  Eq.~(\ref{eq3}). This method has been used successfully, despite of the gain and loss terms that make the underlying system non-conservative \cite{Roy_Raman_2008}. The success of the variational method depends on the choice of a suitable ansatz function for the beam's shape. The method relies on the  assumption that functional form of the beam's shape remains intact in the presence of small perturbations, even though its parameters appearing in the ansatz (amplitude, width, phase, phase-front curvature, etc.) evolve with propagation.

As a first step, we normalize Eq.~(\ref{eq3}) using $\xi = bz$, $r=\rho/w_{s0}$, and $\psi_s= A_s/\sqrt{I_{s0}}$ and rewrite it in the following form,
\begin{align}
    i\frac{\partial \psi_s}{\partial \xi} &+ \frac{\delta}{2}\left(\frac{\partial^2 \psi_s}{\partial r^2} + \frac{1}{r} \frac{\partial \psi_s}{\partial r} \right) - \frac{1}{2\delta} r^2 \psi_s  + \gamma |\psi_s|^2 \psi_s = \notag \\
    &\frac{i}{2} [g(r,\xi)-\alpha_s] \psi_s. \label{eq5}
\end{align}
Here $w_{s0}$ and $I_{s0}$ are the spot size and peak intensity of the input beam and we have introduced two dimensionless parameters as
\begin{equation}
	\delta = (w_g/w_{s0})^2,\qquad \gamma = \omega_s n_2 I_{s0}/(cb),
	\label{eq4b}
\end{equation}
where $w_g = 1/\sqrt{bk_s}$ is the width of the fundamental mode of the GRIN fiber. Typically $w_g$ is close to 5~$\mu$m for GRIN fibers. The gain and loss coefficient are normalized as $g(r,\xi)= G(\rho,z)/b$ and $\alpha_{s}=\alpha_l/b$. 

To implement the variational method, we treat the  term on the right side of Eq.~(\ref{eq5}) as a small perturbation, 
\begin{equation}
    \epsilon = \frac{1}{2}\left[g(r,\xi)-\alpha_s\right] \psi_s,
    \label{eq6}
\end{equation}
where the normalized gain profile has the form $g(r,\xi)=g_0(\xi)\exp(-r^2/r_g^2)$ with $r_g=\rho_g/w_{s0}$. In the case of edge pumping, the peak gain is expected to decrease exponentially because of pump's absorption, i.e., $g_0(\xi)=g_a\exp(-\alpha_g \xi)$, where $g_a$ is the peak gain at the input end of the GRIN fiber and $\alpha_g$ is absorption coefficient of the pump beam. In the case of cladding pumping, the core is pumped uniformally from the cladding side and we set $\alpha_g=0$.

The Lagrangian density $\mathcal{L}_d$ corresponding to Eq.~(\ref{eq5}) has the form \cite{Anderson_Pereira_1999}
\begin{align}
    \mathcal{L}_d =\frac{i}{2} r &\left(\psi_s \partial_{\xi} \psi_{s}^* - \psi_{s}^* \partial_{\xi} \psi_{s} \right) + \frac{\delta}{2} r |\partial_r \psi_s|^2 \nonumber\\
    &- \frac{\gamma}{2} r |\psi_s|^4  + \frac{r^3}{2\delta} |\psi_s|^2 + i r \left(\epsilon \psi_s^* - \epsilon^* \psi_s\right), \label{eq7}
\end{align}
where $\partial_{x} \equiv  \partial/\partial x$. We choose a Gaussian form for our ansatz for $\psi_s$ because the signal is often in the form of a Gaussian beam in practice. It si important to include the curvature of the wavefront and use the form
\begin{equation}
    \psi_s(r,\xi) = \psi_{s0}(\xi) \; {\rm exp} \left[-\frac{r^2}{2r_s^2(\xi)}  + i d_s(\xi) r^2 + i \phi_s(\xi) \right], \label{eq8}
\end{equation}
where the four parameters, $\psi_{s0}, r_s , d_s$ and $\phi_s$, correspond to the beam's amplitude, width, wavefront curvature, and phase, respectively. All of them are allowed to vary with $\xi$. 

Using the preceding ansatz and following the standard Rayleigh--Ritz optimization procedure\cite{Anderson_Pereira_1999}, we obtain the reduced Lagrangian, $L = \int_0^\infty \mathcal{L}_d dr$, by integrating over $r$. The result is found to be
\begin{align}
    L &=& \frac{1}{2} \psi_{s0}^2 r_s^2 \left(\frac{d\phi_s}{d\xi}\right) + \left[2 \delta d_s^2 + \frac{1}{2\delta} + \frac{d d_s}{d\xi}\right] \frac{\psi_{s0}^2 r_s^4}{2} \nonumber \\ && + \frac{\delta}{4} \psi_{s0}^2 - \frac{\gamma}{8} \psi_{s0}^4 r_s^2 + i \int_0^\infty r \left(\epsilon \psi_s^* - \epsilon^* \psi_s \right) dr.
\label{eq9}
\end{align}
Next we use the \textit{Euler-Lagrange} equation, $\partial_\xi (\partial_{X_\xi} L) = \partial_X L$, with $X = \psi_{s0}, r_s, d_s, \phi_s $ and obtain the following four coupled equations for the evolution of the four parameters along the amplifier's length:
\begin{align}
    \frac{d \psi_{s0}}{d \xi} &= - 2 \delta d_s \psi_{s0} -\frac{1}{2}\alpha_s \psi_{s0}+ \frac{g_0 (\xi)}{2} \frac{(1+2\sigma^2)}{(1+\sigma^2)^2} \psi_{s0}, \label{eq10} \\
    \frac{d r_s}{d \xi} &= 2 \delta d_s r_s - \frac{g_0(\xi)}{2} \frac{\sigma^2}{(1+\sigma^2)^2} r_s, \label{eq11} \\
    \frac{d d_s}{d\xi} &= - 2 \delta d_s^2  - \frac{1}{2\delta} + \frac{\delta}{2 r_s^4} - \frac{\gamma}{4} \left( \frac{\psi_{s0}}{r_s} \right)^2, \label{eq12} \\
    \frac{d\phi_s}{d\xi} &= -\frac{\delta}{r_s^2} + \frac{3}{4} \gamma\psi_{s0}^2,    \label{eq13}\\
\end{align}
where $\sigma=r_s/r_g$.  An additional equation is obtained for the beam's power using $P_s=\pi\psi_{s0}^2r_s^2$:
\begin{align}
     \frac{dP_s}{d\xi} &=  -\alpha_s P_s+\frac{g_0(\xi)}{(1+\sigma^2)}P_s. \label{eq13a}
\end{align}
The preceding set of ordinary differential equations (ODEs) can be solved numerically much faster than Eq.~(\ref{eq5}) that governs the evolution of the signal beam. However, accuracy of the resulting solution needs to be checked by solving Eq.~(\ref{eq5}) directly.

It is possible to obtain a single differential equation for the beam width $(r_s)$ under certain approximations. For example, in absence of SPM ($\gamma=0$), $r_s$ satisfies the following differential equation:
\begin{align}
	\frac{d^2 r_s}{d \xi^2} &= - r_s + \frac{\delta^2}{r_s^3} 
    - \left[\frac{2}{(1+\sigma^2)}\frac{d r_s}{d \xi} - \frac{\alpha_g}{2}{r_s} \right] g_0(\xi)\mathcal{F(\sigma)}, \label{eq14}
\end{align}
where $\mathcal{F(\sigma)}=\sigma^2/(1+\sigma^2)^2$. In obtaining this equation, we neglected a higher-order term associated with $g_a$ because it has a negligible contribution. Eq.~\eqref{eq14} offers significant physical insight into the signal beam's evolution in an active GRIN fiber under different pumping schemes. In the absence of gain ($g_0=0$). the width $r_s$ satisfies a simpler equation with the known analytic solution 
\begin{equation}
    r_s(\xi) = [\cos^2(\xi)+ \delta^2\sin^2(\xi)]^{1/2}. 
\end{equation}
It shows that $r_s$ varies periodically with $\xi$ such that the signal beam recovers its initial shape at distances $z=m z_p$ where $m$ is an integer and $z_p=\pi/b$ is the self-imaging period of the GRIN fiber with a typical value of 5~mm. Figure \ref{Fig1}(c) shows schematically the periodic evolution of such signal beam inside the core of a GRIN fiber. For finite values of $g_0$, the last term in Eq.~\eqref{eq14} perturbs the periodicity in such a way that the width keeps oscillating with the same periodicity but deviates from its input value.

\section{Results and Discussion}

In this section we solve the set of ODEs, Eqs.\ (\ref{eq10})--(\ref{eq13}) with the fourth-order Runge--Kutta method and study evolution of four parameters under different pumping conditions. In parallel, we check accuracy of the solution by solving Eq.~(\ref{eq5}) numerically with the standard \textit{Split-Step Fourier} (SSF) method. In both cases, we employ the same values of the parameters given in Table \ref{table:1}. The $\gamma$ and $\delta$ values are calculated for a realistic GRIN fiber designed with $a=50$ $\mu$m and $\Delta=0.01$. For these values, $b=2.8 \times 10^{3}$ m$^{-1}$. The signal power $P_{s0}$ and beam radius $w_{s0}$ at the input end are 100~W and 15 $\mu$m, respectively. The peak gain $g_a$ at the input end depends on the pumping level. We chose a relatively high value to ensure significant amplification over relatively short fiber lengths ($G_a=0.28$ mm$^{-1}$ for $g_a=0.1$).  The initial values used for solving Eqs.\ (\ref{eq10})--(\ref{eq13}) were $\psi_{s0}=1$, $r_{s}=1$, $d_{s}=0$ and $\phi_s=0$.

\begin{table}[tb]
	\begin{center}
		\begin{tabular}{|c|c|c|}
			\hline
			Parameter& Symbols & Values \\
			\hline
			Normalized nonlinear coefficient & $\gamma$ 		& $4 \times 10^{-6}$ 	 \\
			Width ratio & $\delta$ 		& $0.09$ 	 \\
			Normalized gain amplitude& $ g_a$ & $0.1$\\
			Input signal power & $ P_{s0}$ & 100 Watt \\
			Input signal beam width & $ w_{s0}$ & 15 $\mu$m \\
			Linear loss & $ \alpha_l$ &  $2 \times 10^{-3}$ m$^{-1}$ \\
			\hline
		\end{tabular}
		\caption{Parameter values used in the simulations.}
		\label{table:1}
	\end{center}
\end{table}

\subsection{Case I: Cladding Pumping}

\begin{figure} [tb]
	\includegraphics[width=\linewidth]{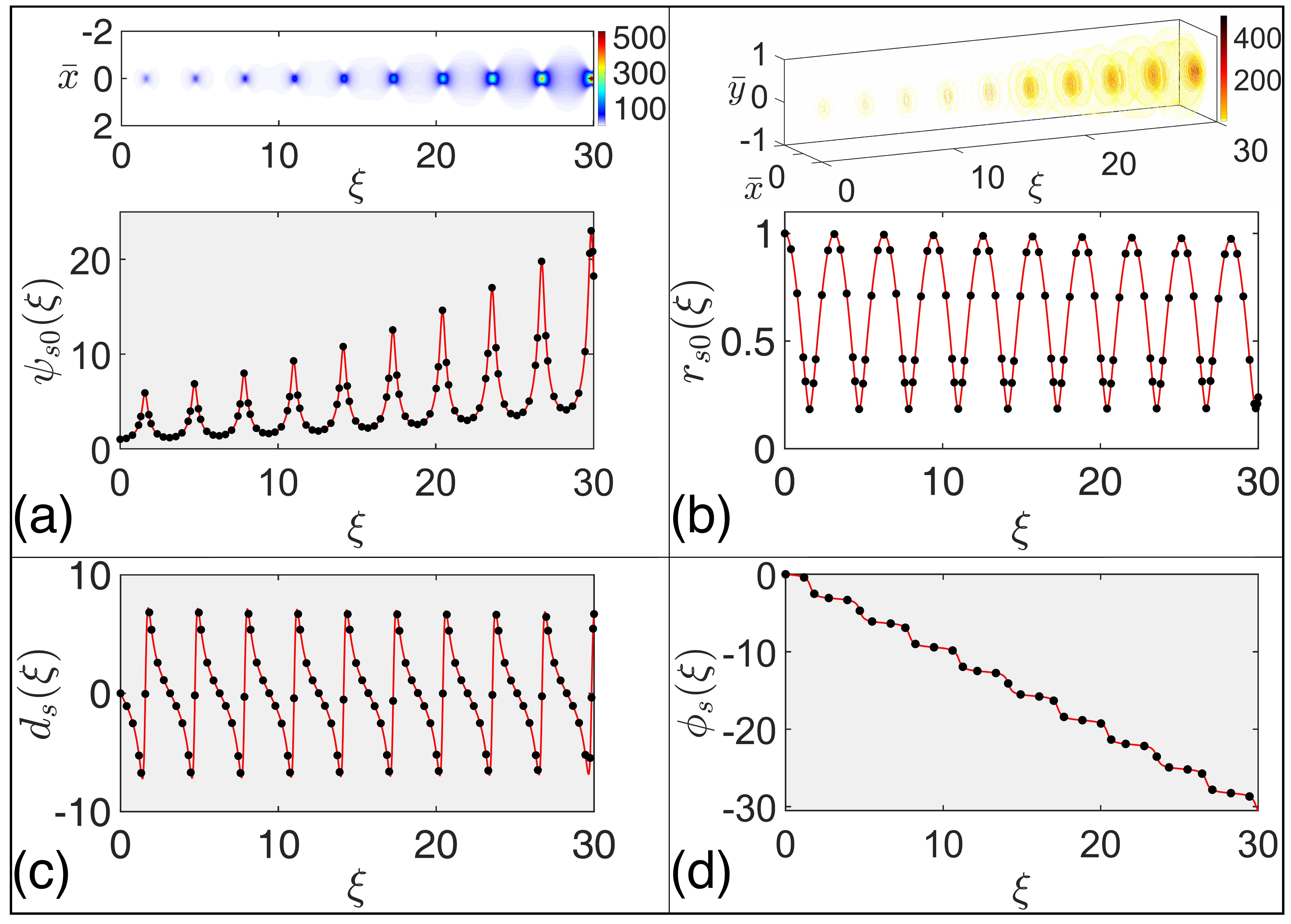}
	\caption{Comparison between the variational (solid lines) and numerical (solid dots) predictions: \textbf{(a)} amplitude, \textbf{(b)} width, \textbf{(c)} phase-front curvature, and \textbf{(d)} phase of the signal beam for $g_0=0.1$ and $\alpha_g=0$. The top inset in \textbf{(a)} shows periodic self-imaging of the signal beam (see text for simulation parameters). The 3D evolution of signal beam is shown as the top inset in \textbf{(b)}.}
	\label{Fig2}
\end{figure}

First we consider the case of a double-clad GRIN fiber that is side-pumped using a relatively wide pump beam. We used $\rho_g=75~\mu$m, which corresponds a full width of 125~$\mu$m for the the input pump beam. In this pumping scheme, the gain change much along the fiber's length and we  can use $g_0(z)=g_a$. The variational results for the beam's amplitude, width, phase-front curvature, and phase are shown in Fig.~\ref{Fig2} as solid lines and compared with full numerical results (soli dots) over a distance that corresponds to nine self-imaging periods. An excellent agreement between the numerical and variational results is evident in Fig.~\ref{Fig2} over this distance. The beam's amplitude increases considerably after each period, its width almost recovers its initial value, indicating an absence of beam narrowing over a short length ($<1$~cm) in the case of clad-pumping.

\subsection{Case II: Edge Pumping}

\begin{figure}[tb]
	\includegraphics[width=\linewidth ]{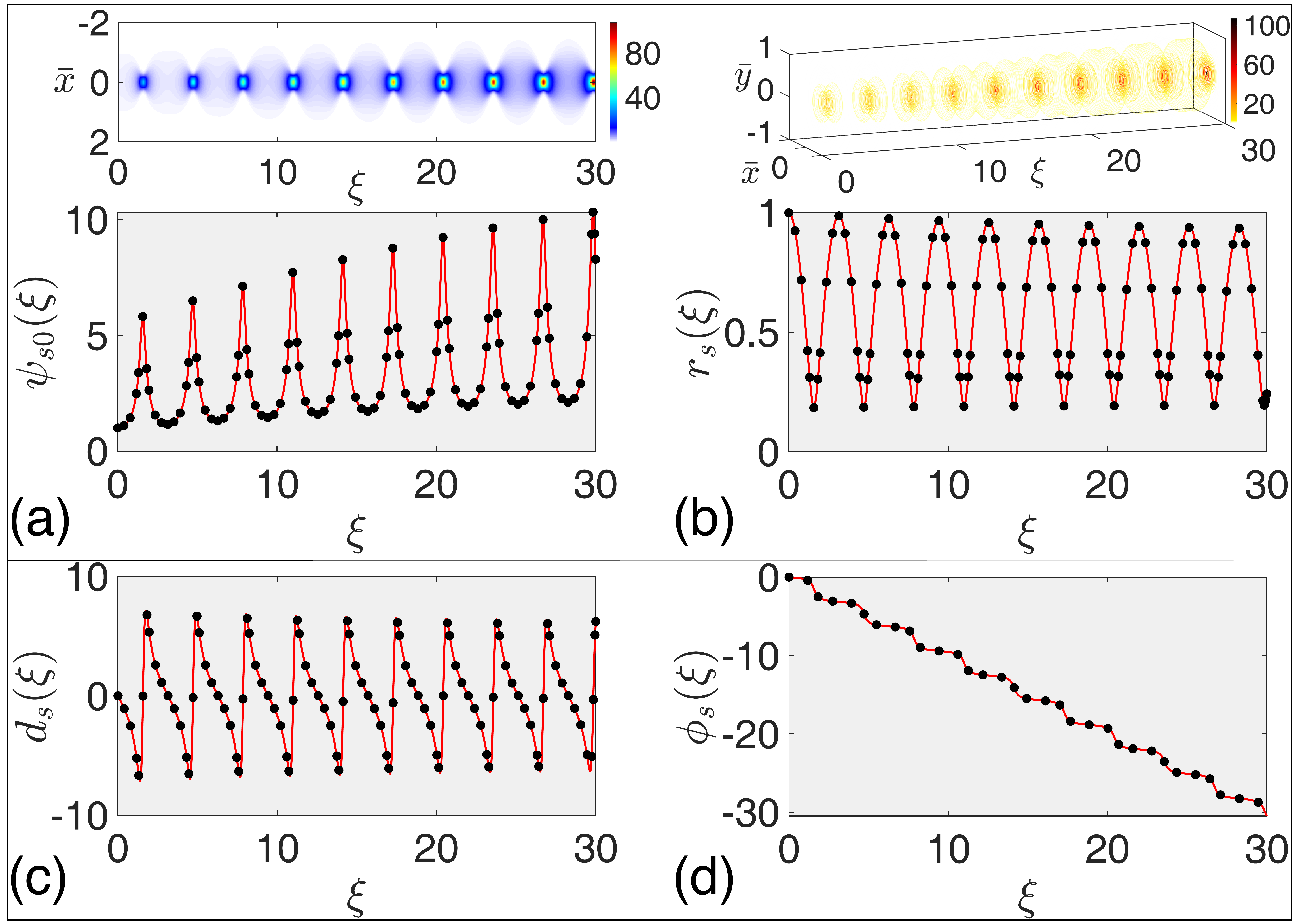}
	\caption{Comparison between the variational (solid lines) and numerical (solid dots) predictions for edge-pumping: \textbf{(a)} amplitude, \textbf{(b)} width, \textbf{(c)} phase-front curvature, and \textbf{(d)} phase of the signal beam. The top inset in \textbf{(a)} shows periodic self-imaging of the signal beam. The peak gain varies with distance $g_0(\xi)=g_a\exp(-\alpha_g \xi)$, with $g_a=0.1$ and $\alpha_g=0.05$. Other parameters are identical to those used in Fig.~\ref{Fig2}.}
	\label{Fig3}
\end{figure}

Next we consider the case edge pumping of a single-clad GRIN fiber that is edge-pumped using a narrower pump beam. We used $\rho_g=30~\mu$m, which corresponds a full width of 50~$\mu$m for the input pump beam. The new feature in this case is that the peak gain at the axis decreases with distance due to the absorption of pump power as $g_0(\xi)=g_a\exp(-\alpha_g \xi)$. As in Fig.~\ref{Fig2}, variational results for the beam's amplitude, width, phase-front curvature, and phase are shown in Fig.~\ref{Fig3} as solid lines and compared with full numerical results (soli dots) over a distance that corresponds to nine self-imaging periods. An excellent agreement is observed again between the two sets of results.  It is evident that less amplification occurs compared to the case of clad-pumping. An interesting feature is that the beam's width does not recover its initial value after each self-imaging values. Its smaller values indicate that some beam narrowing occurs over short distances.

\subsection{Average Behavior}

\begin{figure}[bt!]
	\includegraphics[width=\linewidth ]{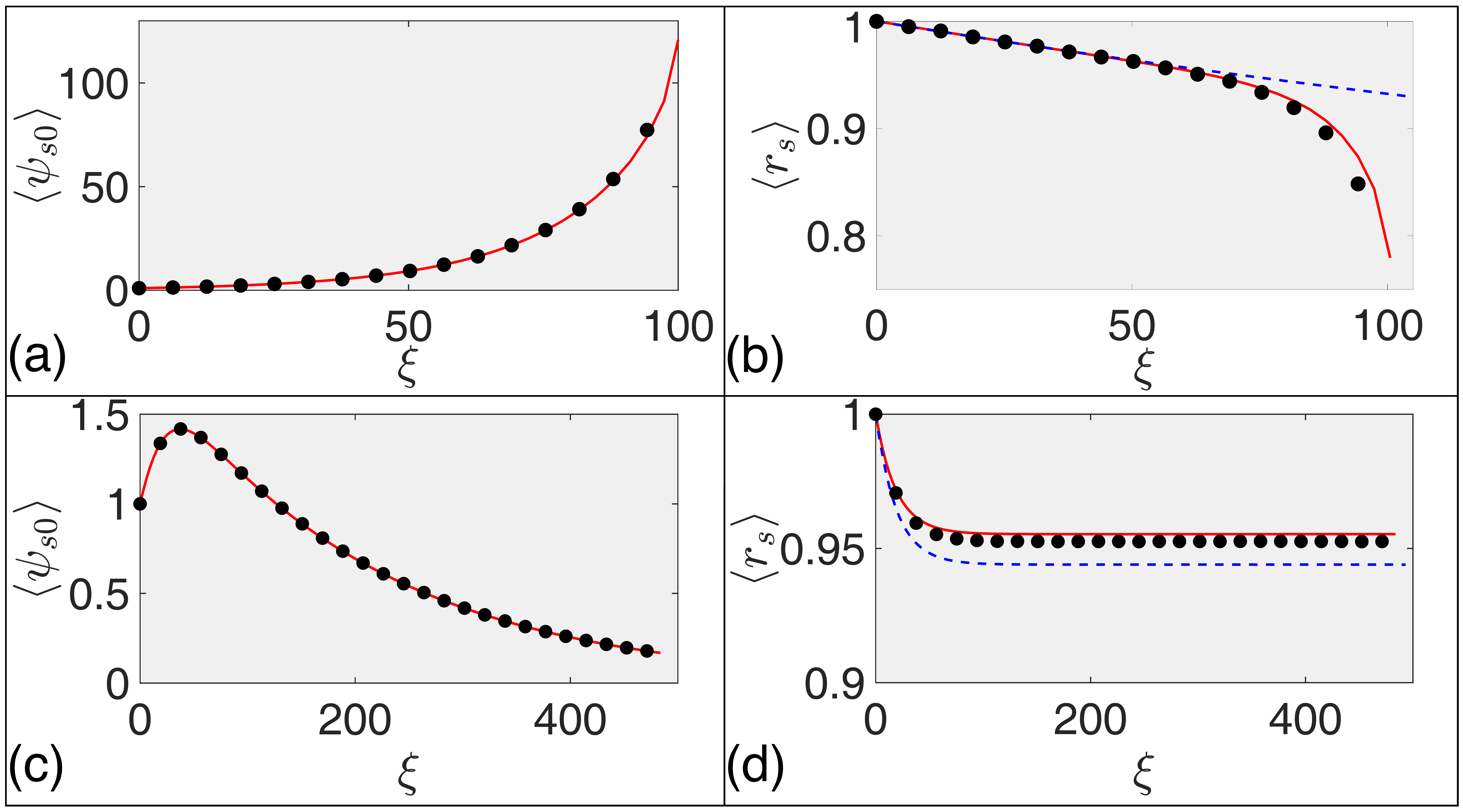}
\caption{Changes with distance of average \textbf{(a,~c)} amplitude $\langle \psi_{s0} \rangle$ and \textbf{(b,~d)} beam width $\langle r_s \rangle$ of the signal beam for the clad-pumping scheme (top row) and edge-pumping scheme (bottom row). The variational results are shown by lines while the solid dots correspond to numerical simulations. Dashed lines in \textbf{(b)} and \textbf{(d)} represent solution of  Eq.~(\ref{eq14}).}
	\label{Fig4}
\end{figure}

Rapid self-imaging oscillations that occur in all GRIN fibers make it harder to draw conclusions about the beam's evolution in real amplifiers whose lengths are long enough that thousands od oscillations can occur. For this reason, we average the two most relevant parameters of the signal beam over such rapid oscillations. We show in Fig.~\ref{Fig4} the averaged values, $\langle \psi_{s0} \rangle$ and $\langle r_s \rangle$, for the two pumping schemes. Top row shows the case of clad-pumping ($\alpha_g=0$), and bottom row shows the case of edge-pumping.  Clearly, the two cases behave quite differently when fiber's length is closer to 1~meter.

In the case of clad-pumping, the amplitude $\langle\psi_{s0}\rangle$ increases monotonically. At the same time, beam narrowing occurs owing to Kerr-induced self-focusing. Indeed, beam collapse seem to occur after 100 periods as the signal beam's power approaches the critical level need for catastrophic self-focusing. This behavior is somewhat artificial because we have used large values of the gain, while ignoring its saturation. Nevertheless, one must be aware of the possibility of beam collapse in high-power GRIN-fiber amplifiers.

In the case of edge-pumping, the amplitude $\langle\psi_{s0}\rangle$ increases initially but begins to decrease after peaking at some distance [see Fig.~\ref{Fig4}(c)]. This decrease is due to exponential reduction in gain with distance. At some point, gain becomes less that the loss, and the signal power begins to decrease. The average beam width $\langle r_s \rangle$ initially decreases but saturates afterwards as seen in part (d). The initial decrease is not due to self-focusing but results from a narrow gain profile. In all cases, variational results (solid lines) are compared with full numerical simulations (solid dots) with excellent agreement between the two. the solution of Eq.~(\ref{eq14}) in parts (b) and (d) with a dashed line. It agrees with the variational and numerical results, except at large distances in part (b). This is sp because Eq.\ (\ref{eq14}) does not include the effect of SPM, which becomes significant when signal is amplified so much the beam collapses owing to self-focusing. 

\begin{figure}[tb!]
	\includegraphics[width=\linewidth ]{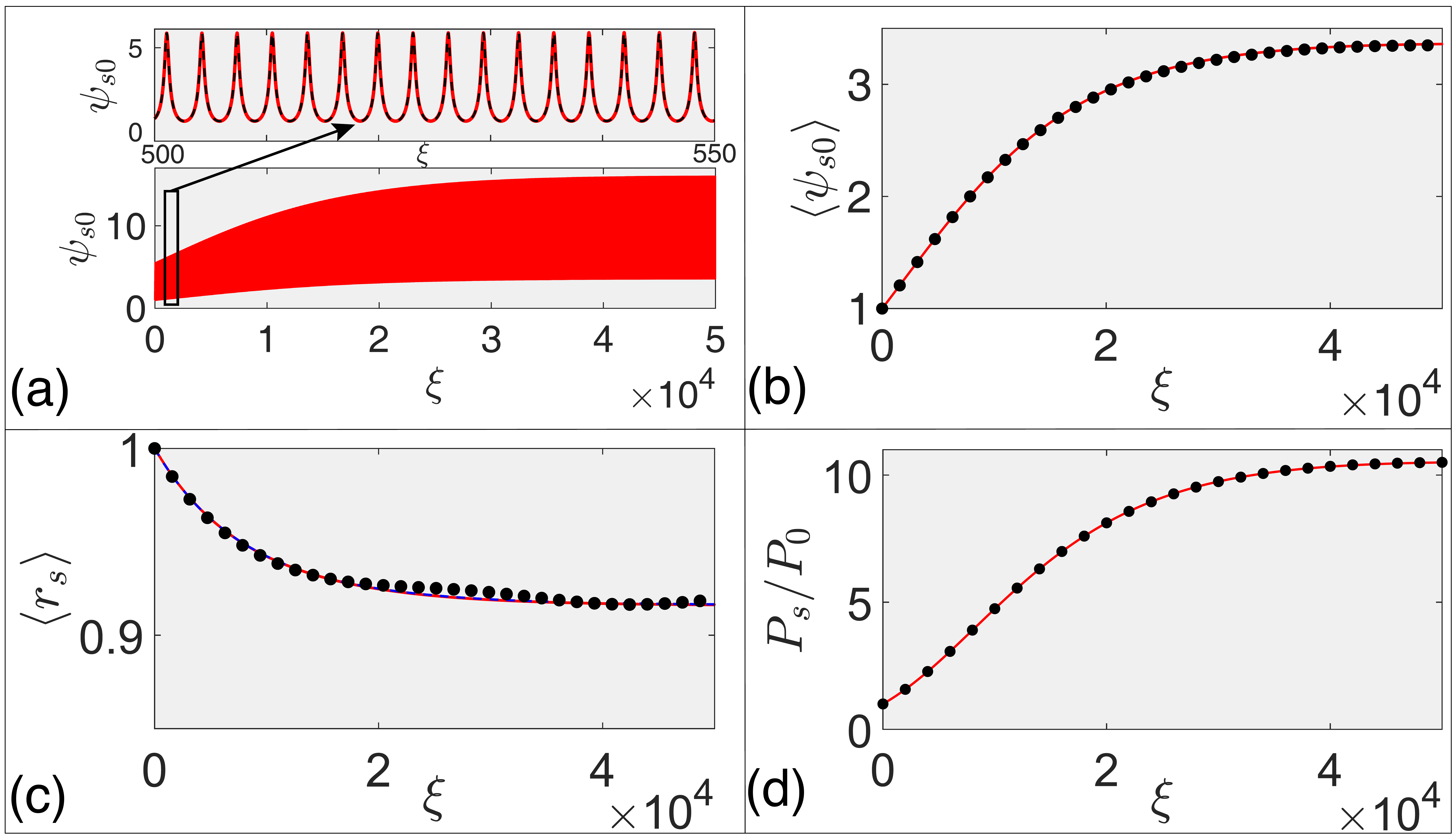}
	\caption{Evolution of signal beam for $z\approx 20$ m in the GRIN fiber with realistic values of gain and loss $G_0=0.8$ m$^{-1}$, $\alpha_l=2 \times 10^{-3}$ m$^{-1}$. The parameters used in the simulation are,  \textbf{(a)} Evolution of the amplitude. In the inset we magnify the periodic evolution of $\psi_{so}$ where full simulation result (dotted line) compared with variational result (solid line).  Variation of average \textbf{(b)} signal amplitude $\langle \psi_{s0} \rangle$ and \textbf{(c)} beam width $\langle r_s \rangle$ with propagation distance. \textbf{(d)} Evolution of the signal power $P_s$ over distance. In all cases the solid dots represent full numerical results and the solid lines correspond to variational results. In plot \textbf{(c)} the dashed line corresponds to the value of  $\langle r_s \rangle$ which is obtained by solving Eq.\eqref{eq14}.  }
	\label{Fig5}
\end{figure}

For our last example, we consider a fiber amplifier in the edge-pumping case with realistic normalized parameters: $g_0=2.82 \times 10^{-4}$, $\alpha_g=1.06\times 10^{-4}$, and $\alpha_s=3.5 \times 10^{-7}$. As the amplifier's length exceeds 10~m in practice, we run our simulations over a distance of about 20~m. Figure \ref{Fig5} shows the evolution of $\psi_{s0}$, $\langle \psi_{s0} \rangle$, $\langle r_s \rangle$, and $P_s/P_0$. In part (a), $\psi_{s0}$ evolves periodically with increasing amplitude owing to the beam's amplification. The inset shows periodic self-imaging on a magnified scale. The average amplitude of the signal beam in part (b) increases first but saturates after some distance because of pump's absorption along the fiber's length. Part (c) shows that the beam's width is reduced first, but its narrowing also saturates when amplification becomes negligible in later sections of the fiber. A dashed line in part (c) shows solution of Eq.\ (\ref{eq14}) for comparison. In plot(d) the evolution of signal's power ($P_s$) is shown with distance. It follows the same trend as $\langle\psi_{s0} \rangle$, i.e., the power increases initially owing to amplification but saturates eventually owing to the loss of pump power at large distances. We compare our variational results (solid line) with full numerical simulation (solid dots) and found them to be in good agreement. It is important to stress that full numerical simulations took more than a day to complete, while variational calculations lasted few minutes on the same computer.

\section{Conclusions}

In this work, a theoretical model is developed based on the variational technique to investigate the evolution of an optical beam inside an optically pumped graded-index fiber amplifier. We consider two pumping schemes used commonly for high-power fiber amplifiers. In the clad-pumping scheme, the use of a relatively wide pump beam results in a nearly uniform gain all along the fiber. In the case of edge pumping, a narrower pump beam propagates together with the signal beam. Its absorption by the dopants provides optical gain that varies both radially and axially along the fiber's length.

The variational technique, having its origin in the Ritz's optimization principle, provides us with a set of coupled ordinary differential equations for the beam's four parameters, its amplitude, width, phase-front curvature, and phase, that change along the amplifier's length Numerical solution of these equations is much faster compared to the solution of the underlying multidimensional nonlinear wave equation. We include in our analysis the nonlinear Kerr term responsible for both SPM and self-focusing. term in the beam dynamics for completeness which eventually leads to beam collapsing due to the self-focusing effect.

We compare the results of the variational and full numerical simulations for the clad-pumping and edge-pumping schemes. In both cases, variational results are found to be in good agreement with time-consuming full numerical simulations. In the case of clad-pumping, the local gain remains nearly constant with distance, and power of the signal beam keeps increasing all along the fiber. In contrast, local gain keeps decreasing along the fiber's length because of pump's absorption by the dopants. As a result, the signal beam is amplified less and less as it propagates inside the fiber. Our variational results predict initial beam narrowing, in  agreement with numerical simulations. We also derive a single equation for the beam's width that can predict amplification-induced narrowing of the signal beam in most cases of practical interest. A major advantage of our variational analysis is that it provides reasonably accurate results much faster compared to the numerical simulations of the multidimensional nonlinear wave equation. In addition, it offers important physical insights and shows what parameters should be controlled to realize spatial narrowing of the amplified beam, a feature that improves the beam's quality at the amplifier's output.

\bibliographystyle{apsrev4-2}
\bibliography{refs,reference}

\begin{thebibliography}{19}%
\makeatletter
\providecommand \@ifxundefined [1]{%
 \@ifx{#1\undefined}
}%
\providecommand \@ifnum [1]{%
 \ifnum #1\expandafter \@firstoftwo
 \else \expandafter \@secondoftwo
 \fi
}%
\providecommand \@ifx [1]{%
 \ifx #1\expandafter \@firstoftwo
 \else \expandafter \@secondoftwo
 \fi
}%
\providecommand \natexlab [1]{#1}%
\providecommand \enquote  [1]{``#1''}%
\providecommand \bibnamefont  [1]{#1}%
\providecommand \bibfnamefont [1]{#1}%
\providecommand \citenamefont [1]{#1}%
\providecommand \href@noop [0]{\@secondoftwo}%
\providecommand \href [0]{\begingroup \@sanitize@url \@href}%
\providecommand \@href[1]{\@@startlink{#1}\@@href}%
\providecommand \@@href[1]{\endgroup#1\@@endlink}%
\providecommand \@sanitize@url [0]{\catcode `\\12\catcode `\$12\catcode
  `\&12\catcode `\#12\catcode `\^12\catcode `\_12\catcode `\%12\relax}%
\providecommand \@@startlink[1]{}%
\providecommand \@@endlink[0]{}%
\providecommand \url  [0]{\begingroup\@sanitize@url \@url }%
\providecommand \@url [1]{\endgroup\@href {#1}{\urlprefix }}%
\providecommand \urlprefix  [0]{URL }%
\providecommand \Eprint [0]{\href }%
\providecommand \doibase [0]{https://doi.org/}%
\providecommand \selectlanguage [0]{\@gobble}%
\providecommand \bibinfo  [0]{\@secondoftwo}%
\providecommand \bibfield  [0]{\@secondoftwo}%
\providecommand \translation [1]{[#1]}%
\providecommand \BibitemOpen [0]{}%
\providecommand \bibitemStop [0]{}%
\providecommand \bibitemNoStop [0]{.\EOS\space}%
\providecommand \EOS [0]{\spacefactor3000\relax}%
\providecommand \BibitemShut  [1]{\csname bibitem#1\endcsname}%
\let\auto@bib@innerbib\@empty
\bibitem [{\citenamefont {Renninger}\ and\ \citenamefont
  {Wise}(2013)}]{Renninger_Optical_2013}%
  \BibitemOpen
  \bibfield  {author} {\bibinfo {author} {\bibfnamefont {W.~H.}\ \bibnamefont
  {Renninger}}\ and\ \bibinfo {author} {\bibfnamefont {F.~W.}\ \bibnamefont
  {Wise}},\ }\href {https://doi.org/10.1038/ncomms2739} {\bibfield  {journal}
  {\bibinfo  {journal} {Nature Communications}\ }\textbf {\bibinfo {volume}
  {4}},\ \bibinfo {pages} {1719} (\bibinfo {year} {2013})}\BibitemShut
  {NoStop}%
\bibitem [{\citenamefont {Wright}\ \emph
  {et~al.}(2015{\natexlab{a}})\citenamefont {Wright}, \citenamefont {Wabnitz},
  \citenamefont {Christodoulides},\ and\ \citenamefont
  {Wise}}]{Wright_Ultrabroadband_2015}%
  \BibitemOpen
  \bibfield  {author} {\bibinfo {author} {\bibfnamefont {L.~G.}\ \bibnamefont
  {Wright}}, \bibinfo {author} {\bibfnamefont {S.}~\bibnamefont {Wabnitz}},
  \bibinfo {author} {\bibfnamefont {D.~N.}\ \bibnamefont {Christodoulides}},\
  and\ \bibinfo {author} {\bibfnamefont {F.~W.}\ \bibnamefont {Wise}},\ }\href
  {https://doi.org/10.1103/physrevlett.115.223902} {\bibfield  {journal}
  {\bibinfo  {journal} {Physical Review Letters}\ }\textbf {\bibinfo {volume}
  {115}},\ \bibinfo {pages} {223902} (\bibinfo {year}
  {2015}{\natexlab{a}})}\BibitemShut {NoStop}%
\bibitem [{\citenamefont {Wright}\ \emph
  {et~al.}(2015{\natexlab{b}})\citenamefont {Wright}, \citenamefont
  {Christodoulides},\ and\ \citenamefont {Wise}}]{Wright_Controllable_2015}%
  \BibitemOpen
  \bibfield  {author} {\bibinfo {author} {\bibfnamefont {L.~G.}\ \bibnamefont
  {Wright}}, \bibinfo {author} {\bibfnamefont {D.~N.}\ \bibnamefont
  {Christodoulides}},\ and\ \bibinfo {author} {\bibfnamefont {F.~W.}\
  \bibnamefont {Wise}},\ }\href {https://doi.org/10.1038/nphoton.2015.61}
  {\bibfield  {journal} {\bibinfo  {journal} {Nature Photonics}\ }\textbf
  {\bibinfo {volume} {9}},\ \bibinfo {pages} {306} (\bibinfo {year}
  {2015}{\natexlab{b}})}\BibitemShut {NoStop}%
\bibitem [{\citenamefont {Wright}\ \emph {et~al.}(2016)\citenamefont {Wright},
  \citenamefont {Liu}, \citenamefont {Nolan}, \citenamefont {Li}, \citenamefont
  {Christodoulides},\ and\ \citenamefont {Wise}}]{Wright_Self_2016}%
  \BibitemOpen
  \bibfield  {author} {\bibinfo {author} {\bibfnamefont {L.~G.}\ \bibnamefont
  {Wright}}, \bibinfo {author} {\bibfnamefont {Z.}~\bibnamefont {Liu}},
  \bibinfo {author} {\bibfnamefont {D.~A.}\ \bibnamefont {Nolan}}, \bibinfo
  {author} {\bibfnamefont {M.-J.}\ \bibnamefont {Li}}, \bibinfo {author}
  {\bibfnamefont {D.~N.}\ \bibnamefont {Christodoulides}},\ and\ \bibinfo
  {author} {\bibfnamefont {F.~W.}\ \bibnamefont {Wise}},\ }\href
  {https://doi.org/10.1038/nphoton.2016.227} {\bibfield  {journal} {\bibinfo
  {journal} {Nature Photonics}\ }\textbf {\bibinfo {volume} {10}},\ \bibinfo
  {pages} {771} (\bibinfo {year} {2016})}\BibitemShut {NoStop}%
\bibitem [{\citenamefont {Wright}\ \emph {et~al.}(2017)\citenamefont {Wright},
  \citenamefont {Christodoulides},\ and\ \citenamefont
  {Wise}}]{Wright_Spatiotemporal_2017}%
  \BibitemOpen
  \bibfield  {author} {\bibinfo {author} {\bibfnamefont {L.~G.}\ \bibnamefont
  {Wright}}, \bibinfo {author} {\bibfnamefont {D.~N.}\ \bibnamefont
  {Christodoulides}},\ and\ \bibinfo {author} {\bibfnamefont {F.~W.}\
  \bibnamefont {Wise}},\ }\href {https://doi.org/10.1126/science.aao0831}
  {\bibfield  {journal} {\bibinfo  {journal} {Science}\ }\textbf {\bibinfo
  {volume} {358}},\ \bibinfo {pages} {94} (\bibinfo {year} {2017})}\BibitemShut
  {NoStop}%
\bibitem [{\citenamefont {Qin}\ \emph {et~al.}(2018)\citenamefont {Qin},
  \citenamefont {Xiao}, \citenamefont {Wang},\ and\ \citenamefont
  {Yang}}]{Qin_Observation_2018}%
  \BibitemOpen
  \bibfield  {author} {\bibinfo {author} {\bibfnamefont {H.}~\bibnamefont
  {Qin}}, \bibinfo {author} {\bibfnamefont {X.}~\bibnamefont {Xiao}}, \bibinfo
  {author} {\bibfnamefont {P.}~\bibnamefont {Wang}},\ and\ \bibinfo {author}
  {\bibfnamefont {C.}~\bibnamefont {Yang}},\ }\href
  {https://doi.org/10.1364/ol.43.001982} {\bibfield  {journal} {\bibinfo
  {journal} {Optics Letters}\ }\textbf {\bibinfo {volume} {43}},\ \bibinfo
  {pages} {1982} (\bibinfo {year} {2018})}\BibitemShut {NoStop}%
\bibitem [{\citenamefont {Lopez-Galmiche}\ \emph {et~al.}(2016)\citenamefont
  {Lopez-Galmiche}, \citenamefont {Eznaveh}, \citenamefont {Eftekhar},
  \citenamefont {Lopez}, \citenamefont {Wright}, \citenamefont {Wise},
  \citenamefont {Christodoulides},\ and\ \citenamefont
  {Correa}}]{LopezGalmiche_Visible_2016}%
  \BibitemOpen
  \bibfield  {author} {\bibinfo {author} {\bibfnamefont {G.}~\bibnamefont
  {Lopez-Galmiche}}, \bibinfo {author} {\bibfnamefont {Z.~S.}\ \bibnamefont
  {Eznaveh}}, \bibinfo {author} {\bibfnamefont {M.~A.}\ \bibnamefont
  {Eftekhar}}, \bibinfo {author} {\bibfnamefont {J.~A.}\ \bibnamefont {Lopez}},
  \bibinfo {author} {\bibfnamefont {L.~G.}\ \bibnamefont {Wright}}, \bibinfo
  {author} {\bibfnamefont {F.}~\bibnamefont {Wise}}, \bibinfo {author}
  {\bibfnamefont {D.}~\bibnamefont {Christodoulides}},\ and\ \bibinfo {author}
  {\bibfnamefont {R.~A.}\ \bibnamefont {Correa}},\ }\href
  {https://doi.org/10.1364/ol.41.002553} {\bibfield  {journal} {\bibinfo
  {journal} {Optics Letters}\ }\textbf {\bibinfo {volume} {41}},\ \bibinfo
  {pages} {2553} (\bibinfo {year} {2016})}\BibitemShut {NoStop}%
\bibitem [{\citenamefont {Eslami}\ \emph {et~al.}(2022)\citenamefont {Eslami},
  \citenamefont {Salmela}, \citenamefont {Filipkowski}, \citenamefont {Pysz},
  \citenamefont {Klimczak}, \citenamefont {Buczynski}, \citenamefont {Dudley},\
  and\ \citenamefont {Genty}}]{Eslami_Two_2022}%
  \BibitemOpen
  \bibfield  {author} {\bibinfo {author} {\bibfnamefont {Z.}~\bibnamefont
  {Eslami}}, \bibinfo {author} {\bibfnamefont {L.}~\bibnamefont {Salmela}},
  \bibinfo {author} {\bibfnamefont {A.}~\bibnamefont {Filipkowski}}, \bibinfo
  {author} {\bibfnamefont {D.}~\bibnamefont {Pysz}}, \bibinfo {author}
  {\bibfnamefont {M.}~\bibnamefont {Klimczak}}, \bibinfo {author}
  {\bibfnamefont {R.}~\bibnamefont {Buczynski}}, \bibinfo {author}
  {\bibfnamefont {J.~M.}\ \bibnamefont {Dudley}},\ and\ \bibinfo {author}
  {\bibfnamefont {G.}~\bibnamefont {Genty}},\ }\href
  {https://doi.org/10.1038/s41467-022-29776-6} {\bibfield  {journal} {\bibinfo
  {journal} {Nature Communications}\ }\textbf {\bibinfo {volume} {13}},\
  \bibinfo {pages} {2126} (\bibinfo {year} {2022})}\BibitemShut {NoStop}%
\bibitem [{\citenamefont {Guenard}\ \emph {et~al.}(2017)\citenamefont
  {Guenard}, \citenamefont {Krupa}, \citenamefont {Dupiol}, \citenamefont
  {Fabert}, \citenamefont {Bendahmane}, \citenamefont {Kermene}, \citenamefont
  {Desfarges-Berthelemot}, \citenamefont {Auguste}, \citenamefont {Tonello},
  \citenamefont {Barth{\'{e}}l{\'{e}}my}, \citenamefont {Millot}, \citenamefont
  {Wabnitz},\ and\ \citenamefont {Couderc}}]{Guenard_Kerr_2017}%
  \BibitemOpen
  \bibfield  {author} {\bibinfo {author} {\bibfnamefont {R.}~\bibnamefont
  {Guenard}}, \bibinfo {author} {\bibfnamefont {K.}~\bibnamefont {Krupa}},
  \bibinfo {author} {\bibfnamefont {R.}~\bibnamefont {Dupiol}}, \bibinfo
  {author} {\bibfnamefont {M.}~\bibnamefont {Fabert}}, \bibinfo {author}
  {\bibfnamefont {A.}~\bibnamefont {Bendahmane}}, \bibinfo {author}
  {\bibfnamefont {V.}~\bibnamefont {Kermene}}, \bibinfo {author} {\bibfnamefont
  {A.}~\bibnamefont {Desfarges-Berthelemot}}, \bibinfo {author} {\bibfnamefont
  {J.~L.}\ \bibnamefont {Auguste}}, \bibinfo {author} {\bibfnamefont
  {A.}~\bibnamefont {Tonello}}, \bibinfo {author} {\bibfnamefont
  {A.}~\bibnamefont {Barth{\'{e}}l{\'{e}}my}}, \bibinfo {author} {\bibfnamefont
  {G.}~\bibnamefont {Millot}}, \bibinfo {author} {\bibfnamefont
  {S.}~\bibnamefont {Wabnitz}},\ and\ \bibinfo {author} {\bibfnamefont
  {V.}~\bibnamefont {Couderc}},\ }\href {https://doi.org/10.1364/oe.25.004783}
  {\bibfield  {journal} {\bibinfo  {journal} {Optics Express}\ }\textbf
  {\bibinfo {volume} {25}},\ \bibinfo {pages} {4783} (\bibinfo {year}
  {2017})}\BibitemShut {NoStop}%
\bibitem [{\citenamefont {Jima}\ \emph {et~al.}(2022)\citenamefont {Jima},
  \citenamefont {Tonello}, \citenamefont {Niang}, \citenamefont {Mansuryan},
  \citenamefont {Krupa}, \citenamefont {Modotto}, \citenamefont {Cucinotta},
  \citenamefont {Couderc},\ and\ \citenamefont
  {Wabnitz}}]{Jima_Numerical_2022}%
  \BibitemOpen
  \bibfield  {author} {\bibinfo {author} {\bibfnamefont {M.~A.}\ \bibnamefont
  {Jima}}, \bibinfo {author} {\bibfnamefont {A.}~\bibnamefont {Tonello}},
  \bibinfo {author} {\bibfnamefont {A.}~\bibnamefont {Niang}}, \bibinfo
  {author} {\bibfnamefont {T.}~\bibnamefont {Mansuryan}}, \bibinfo {author}
  {\bibfnamefont {K.}~\bibnamefont {Krupa}}, \bibinfo {author} {\bibfnamefont
  {D.}~\bibnamefont {Modotto}}, \bibinfo {author} {\bibfnamefont
  {A.}~\bibnamefont {Cucinotta}}, \bibinfo {author} {\bibfnamefont
  {V.}~\bibnamefont {Couderc}},\ and\ \bibinfo {author} {\bibfnamefont
  {S.}~\bibnamefont {Wabnitz}},\ }\href {https://doi.org/10.1364/josab.463473}
  {\bibfield  {journal} {\bibinfo  {journal} {Journal of the Optical Society of
  America B}\ }\textbf {\bibinfo {volume} {39}},\ \bibinfo {pages} {2172}
  (\bibinfo {year} {2022})}\BibitemShut {NoStop}%
\bibitem [{\citenamefont {Mangini}\ \emph {et~al.}(2022)\citenamefont
  {Mangini}, \citenamefont {Gervaziev}, \citenamefont {Ferraro}, \citenamefont
  {Kharenko}, \citenamefont {Zitelli}, \citenamefont {Sun}, \citenamefont
  {Couderc}, \citenamefont {Podivilov}, \citenamefont {Babin},\ and\
  \citenamefont {Wabnitz}}]{Mangini_Statistical_2022}%
  \BibitemOpen
  \bibfield  {author} {\bibinfo {author} {\bibfnamefont {F.}~\bibnamefont
  {Mangini}}, \bibinfo {author} {\bibfnamefont {M.}~\bibnamefont {Gervaziev}},
  \bibinfo {author} {\bibfnamefont {M.}~\bibnamefont {Ferraro}}, \bibinfo
  {author} {\bibfnamefont {D.~S.}\ \bibnamefont {Kharenko}}, \bibinfo {author}
  {\bibfnamefont {M.}~\bibnamefont {Zitelli}}, \bibinfo {author} {\bibfnamefont
  {Y.}~\bibnamefont {Sun}}, \bibinfo {author} {\bibfnamefont {V.}~\bibnamefont
  {Couderc}}, \bibinfo {author} {\bibfnamefont {E.~V.}\ \bibnamefont
  {Podivilov}}, \bibinfo {author} {\bibfnamefont {S.~A.}\ \bibnamefont
  {Babin}},\ and\ \bibinfo {author} {\bibfnamefont {S.}~\bibnamefont
  {Wabnitz}},\ }\href {https://doi.org/10.1364/oe.449187} {\bibfield  {journal}
  {\bibinfo  {journal} {Optics Express}\ }\textbf {\bibinfo {volume} {30}},\
  \bibinfo {pages} {10850} (\bibinfo {year} {2022})}\BibitemShut {NoStop}%
\bibitem [{\citenamefont {Haig}\ \emph {et~al.}(2023)\citenamefont {Haig},
  \citenamefont {Bender}, \citenamefont {Chen}, \citenamefont {Dhar},
  \citenamefont {Choudhury}, \citenamefont {Sen}, \citenamefont
  {Christodoulides},\ and\ \citenamefont {Wise}}]{Haig_Gain_2023}%
  \BibitemOpen
  \bibfield  {author} {\bibinfo {author} {\bibfnamefont {H.}~\bibnamefont
  {Haig}}, \bibinfo {author} {\bibfnamefont {N.}~\bibnamefont {Bender}},
  \bibinfo {author} {\bibfnamefont {Y.-H.}\ \bibnamefont {Chen}}, \bibinfo
  {author} {\bibfnamefont {A.}~\bibnamefont {Dhar}}, \bibinfo {author}
  {\bibfnamefont {N.}~\bibnamefont {Choudhury}}, \bibinfo {author}
  {\bibfnamefont {R.}~\bibnamefont {Sen}}, \bibinfo {author} {\bibfnamefont
  {D.~N.}\ \bibnamefont {Christodoulides}},\ and\ \bibinfo {author}
  {\bibfnamefont {F.}~\bibnamefont {Wise}},\ }\href
  {https://doi.org/10.1364/josab.492262} {\bibfield  {journal} {\bibinfo
  {journal} {Journal of the Optical Society of America B}\ }\textbf {\bibinfo
  {volume} {40}},\ \bibinfo {pages} {1510} (\bibinfo {year}
  {2023})}\BibitemShut {NoStop}%
\bibitem [{\citenamefont {Sidelnikov}\ \emph {et~al.}(2022)\citenamefont
  {Sidelnikov}, \citenamefont {Podivilov}, \citenamefont {Fedoruk},
  \citenamefont {Kuznetsov}, \citenamefont {Wabnitz},\ and\ \citenamefont
  {Babin}}]{Sidelnikov_Mechanism_2022}%
  \BibitemOpen
  \bibfield  {author} {\bibinfo {author} {\bibfnamefont {O.~S.}\ \bibnamefont
  {Sidelnikov}}, \bibinfo {author} {\bibfnamefont {E.~V.}\ \bibnamefont
  {Podivilov}}, \bibinfo {author} {\bibfnamefont {M.~P.}\ \bibnamefont
  {Fedoruk}}, \bibinfo {author} {\bibfnamefont {A.~G.}\ \bibnamefont
  {Kuznetsov}}, \bibinfo {author} {\bibfnamefont {S.}~\bibnamefont {Wabnitz}},\
  and\ \bibinfo {author} {\bibfnamefont {S.~A.}\ \bibnamefont {Babin}},\ }\href
  {https://doi.org/10.1364/oe.449773} {\bibfield  {journal} {\bibinfo
  {journal} {Optics Express}\ }\textbf {\bibinfo {volume} {30}},\ \bibinfo
  {pages} {8212} (\bibinfo {year} {2022})}\BibitemShut {NoStop}%
\bibitem [{\citenamefont {Agrawal}(2023)}]{Agrawal_Spatial_2023}%
  \BibitemOpen
  \bibfield  {author} {\bibinfo {author} {\bibfnamefont {G.~P.}\ \bibnamefont
  {Agrawal}},\ }\href {https://doi.org/10.1364/josab.482730} {\bibfield
  {journal} {\bibinfo  {journal} {Journal of the Optical Society of America B}\
  }\textbf {\bibinfo {volume} {40}},\ \bibinfo {pages} {715} (\bibinfo {year}
  {2023})}\BibitemShut {NoStop}%
\bibitem [{\citenamefont {Saleh}\ \emph {et~al.}(1990)\citenamefont {Saleh},
  \citenamefont {Jopson}, \citenamefont {Evankow},\ and\ \citenamefont
  {Aspell}}]{Saleh_Modeling_1990}%
  \BibitemOpen
  \bibfield  {author} {\bibinfo {author} {\bibfnamefont {A.}~\bibnamefont
  {Saleh}}, \bibinfo {author} {\bibfnamefont {R.}~\bibnamefont {Jopson}},
  \bibinfo {author} {\bibfnamefont {J.}~\bibnamefont {Evankow}},\ and\ \bibinfo
  {author} {\bibfnamefont {J.}~\bibnamefont {Aspell}},\ }\href
  {https://doi.org/10.1109/68.60769} {\bibfield  {journal} {\bibinfo  {journal}
  {IEEE Photonics Technology Letters}\ }\textbf {\bibinfo {volume} {2}},\
  \bibinfo {pages} {714} (\bibinfo {year} {1990})}\BibitemShut {NoStop}%
\bibitem [{\citenamefont {Chen}\ \emph {et~al.}(2023)\citenamefont {Chen},
  \citenamefont {Haig}, \citenamefont {Wu}, \citenamefont {Ziegler},\ and\
  \citenamefont {Wise}}]{Chen:23}%
  \BibitemOpen
  \bibfield  {author} {\bibinfo {author} {\bibfnamefont {Y.-H.}\ \bibnamefont
  {Chen}}, \bibinfo {author} {\bibfnamefont {H.}~\bibnamefont {Haig}}, \bibinfo
  {author} {\bibfnamefont {Y.}~\bibnamefont {Wu}}, \bibinfo {author}
  {\bibfnamefont {Z.}~\bibnamefont {Ziegler}},\ and\ \bibinfo {author}
  {\bibfnamefont {F.}~\bibnamefont {Wise}},\ }\href
  {https://doi.org/10.1364/JOSAB.500586} {\bibfield  {journal} {\bibinfo
  {journal} {J. Opt. Soc. Am. B}\ }\textbf {\bibinfo {volume} {40}},\ \bibinfo
  {pages} {2633} (\bibinfo {year} {2023})}\BibitemShut {NoStop}%
\bibitem [{\citenamefont {Anderson}(1983)}]{Anderson_Variational_1983}%
  \BibitemOpen
  \bibfield  {author} {\bibinfo {author} {\bibfnamefont {D.}~\bibnamefont
  {Anderson}},\ }\href {https://doi.org/10.1103/physreva.27.3135} {\bibfield
  {journal} {\bibinfo  {journal} {Physical Review A}\ }\textbf {\bibinfo
  {volume} {27}},\ \bibinfo {pages} {3135} (\bibinfo {year}
  {1983})}\BibitemShut {NoStop}%
\bibitem [{\citenamefont {Roy}\ \emph {et~al.}(2008)\citenamefont {Roy},
  \citenamefont {Bhadra},\ and\ \citenamefont {Agrawal}}]{Roy_Raman_2008}%
  \BibitemOpen
  \bibfield  {author} {\bibinfo {author} {\bibfnamefont {S.}~\bibnamefont
  {Roy}}, \bibinfo {author} {\bibfnamefont {S.~K.}\ \bibnamefont {Bhadra}},\
  and\ \bibinfo {author} {\bibfnamefont {G.~P.}\ \bibnamefont {Agrawal}},\
  }\href {https://doi.org/10.1364/josab.26.000017} {\bibfield  {journal}
  {\bibinfo  {journal} {Journal of the Optical Society of America B}\ }\textbf
  {\bibinfo {volume} {26}},\ \bibinfo {pages} {17} (\bibinfo {year}
  {2008})}\BibitemShut {NoStop}%
\bibitem [{\citenamefont {Anderson}\ \emph {et~al.}(1999)\citenamefont
  {Anderson}, \citenamefont {Cattani},\ and\ \citenamefont
  {Lisak}}]{Anderson_Pereira_1999}%
  \BibitemOpen
  \bibfield  {author} {\bibinfo {author} {\bibfnamefont {D.}~\bibnamefont
  {Anderson}}, \bibinfo {author} {\bibfnamefont {F.}~\bibnamefont {Cattani}},\
  and\ \bibinfo {author} {\bibfnamefont {M.}~\bibnamefont {Lisak}},\ }\href
  {https://doi.org/10.1238/physica.topical.082a00032} {\bibfield  {journal}
  {\bibinfo  {journal} {Physica Scripta}\ }\textbf {\bibinfo {volume} {T82}},\
  \bibinfo {pages} {32} (\bibinfo {year} {1999})}\BibitemShut {NoStop}%
\end{thebibliography}%

\end{document}